\documentclass[fleqn,usenatbib]{mnras}

\usepackage[T1]{fontenc}
\usepackage{ae,aecompl}

\def\specchar#1{{\sc #1}}

\def\HeI{\mbox{He \specchar{i}}}
\def\CaIIH{\mbox{Ca\,\specchar{ii}\,\,H}}       
\def\CaII{\mbox{Ca\,\specchar{ii}}}
\def\hazel{\textsc{Hazel}}

\def\eg{{e.g.}}

\usepackage{natbib,graphicx}
\usepackage{amsmath}

\usepackage{color}

\usepackage{hyperref}
\hypersetup{
    final=true,
    pageanchor=true,
    colorlinks=true,
    breaklinks=true,
    linkcolor=blue,
    citecolor=blue,
    urlcolor=blue,
    pdfpagemode=UseNone,
    pdftitle={Synthetic polarimetric spectra from stellar prominences},
    pdfauthor={T. Felipe et al.},
    pdfsubject={Solar Physics},
    pdfkeywords={Methods: numerical, Techniques: polarimetric, Line: profiles, Stars: coronae, Stars: activity}}

\title[Synthetic polarimetric spectra from stellar prominences]{Synthetic polarimetric spectra from stellar prominences}

\author[T. Felipe et al.]{
T. Felipe,$^{1,2}$\thanks{E-mail: tobias@iac.es}
M. Mart\'inez Gonz\'alez,$^{1,2}$
and A. Asensio Ramos$^{1,2}$
\\
$^{1}$Instituto de Astrof\'{\i}sica de Canarias, 38205,
C/ V\'{\i}a L{\'a}ctea, s/n, La Laguna, Tenerife, Spain\\
$^{2}$Departamento de Astrof\'{\i}sica, Universidad de La Laguna, 38205, La Laguna, Tenerife, Spain
}

\date{Accepted XXX. Received YYY; in original form ZZZ}

\pubyear{2016}

\begin{document}
\label{firstpage}
\pagerange{\pageref{firstpage}--\pageref{lastpage}}
\maketitle

\begin{abstract}
Stellar prominences detected in rapidly rotating stars serve as probes of the magnetism in the corona of cool stars. We have synthesized the temporal evolution of the Stokes profiles generated in the \HeI\ 10830 and 5876 \AA\ triplets during the rotation of a prominence around a star. The synthesis was performed with the \hazel\ code using a cloud model in which the prominence is characterized by a slab located at a fixed latitude and height. It accounts for the scattering polarization and Zeeman and Hanle effects. Several cases with different prominence magnetic field strengths and orientations have been analyzed. The results show an emission feature that drifts across the profile while the prominence is out of the stellar disk. When the prominence eclipses the star, the intensity profile shows an absorption. The scattering induced by the prominence generates linear polarization signals in Stokes $Q$ and $U$ profiles, which are modified by the Hanle effect when a magnetic field is present. Due to the Zeeman effect, Stokes $V$ profiles show a signal with very low amplitude when the magnetic field along the line-of-sight is different from zero. The estimated linear polarization signals could potentially be detected with the future spectropolarimeter MIRADAS, to be attached to GTC telescope.

\end{abstract}

\begin{keywords}
Methods: numerical -- Techniques: polarimetric -- Line: profiles -- Stars: coronae -- Stars: activity
\end{keywords}


\section{Introduction}

The detection of clouds of material on stellar coronae over the last thirty years has revealed that the corona of late-type stars is a complex media with non-homogeneous density and temperature distributions. These clouds were first reported by \citet{CollierCameron+Robinson1989a}, who identified transient absorption features passing through the Doppler broadened stellar spectrum during the course of 1-2 hours. Their investigation analyzed H$\alpha$ on the rapidly rotating K0 dwarf AB Doradus (AB Dor). This star has been the subject under study on many subsequent works \citep[\eg,][]{CollierCameron+Robinson1989b, Donati+CollierCameron1997, Donati+etal1999}, which have been focused on the analysis of the strong Balmer lines, mostly H$\alpha$ and H$\beta$.

These condensations consist of dense cool material surrounded by the hot stellar coronal atmosphere. The strong magnetic field of the active stars where they are observed suggests that magnetic field plays a major role in supporting these structures against gravity and isolating them from the ambient corona. Their physical properties present some similarities with solar prominences, since their temperature is low enough and their density sufficiently high to contain significant amounts of neutral hydrogen. Thus, they are usually referred as ``stellar prominences'' by analogy to those observed on the Sun. 

Solar prominences are called filaments when they are observed on the solar disk as dark elongated structures. Photons coming from the solar surface behind the prominence are scattered isotropically at the dense prominence plasma, reducing the photons along the line-of-sight (LOS) and producing an absorption feature. This process is equivalent to that generating the absorption transits which led to the detection of the stellar prominences discussed in the previous paragraphs. Solar prominences can also be seen as emission features when they are at off-limb locations, since photons originating from the surface can be scattered into the LOS. Similarly, stellar prominences can also produce emission features when they are out of the stellar disk. H$\alpha$ emission has been identified in two G dwarfs \citep{Donati+etal2000, Barnes+etal2001} and it was interpreted as an inhomogeneous ring of prominence material. \citet{Dunstone+etal2006b} detected emission loops in H$\alpha$ from the rapidly rotating star Speedy Mic and associated them with the absorption features from prominences transiting the stellar disk. 

Stellar prominences show significant differences with solar prominences. The most remarkable one is their height above the stellar surface. Generally, stellar prominences are found as absorption features whose radial velocity relative to the underlying stellar spectrum increases with time during their transit. While active regions located at the stellar surface can show absorption features in photospheric lines with a drift rate given by the surface rotation of the star, the rate at which stellar prominences drift across the profile is much faster. It depends on the time that the prominence spends in front of the star and, thus, it is directly proportional to the distance of the condensation from the stellar rotation axis. They have been detected at radial distances from the axis of the star ranging from 2 to 8 stellar radii ($R_*$), although their distribution shows a preference for cylindrical radial distances close to the Keplerian co-rotation radius (\eg, $2.7R_*$ in the case of AB Dor). 

Out of the co-rotation radius the centrifugal force is not balanced by gravity, and an additional force is required to compensate it and allow the formation of a stable prominence. This force is expected to be the magnetic tension, in such a way that the curvature at the top of coronal loops can hold the condensed material. However, this interpretation presents some challenges. X-ray and EUV measurements suggest that the closed coronal field must be relatively compact \citep{Hussain+etal2005}. In addition, the surface magnetic topology of those stars is very complex, with regions of opposite polarity located all around the star \citep{Donati+CollierCameron1997, Donati+etal1999}. This multipolar field must drop fast with height. These two results indicate that a closed corona unlikely extends to the large radial distances where prominences have been detected. For example, \citet{Hussain+etal2007} found that the X-ray corona from AB Dor must be concentrated at a height of $0.3-0.4R_*$, well below its co-rotation radius. \citet{Jardine+vanBallegooijen2005} proposed an open field coronal model where prominences are supported by closed loops above coronal helmet streamers formed from the reconnection of opposite directed wind-bearing field lines. According to this model, stellar prominences should be found above neutral polarity lines, like on the Sun.

Inferring the magnetic field strength and orientation from stellar prominences could provide compelling insights regarding the magnetic topology of stellar coronae. These studies would benefit from the analysis of the full polarization signals, including the four Stokes parameters $I$, $Q$, $U$, and $V$. However, due to technical limitations, until today all the studies focused on stellar prominences have been restricted to the inspection of intensity spectra. Magnetism on stellar surfaces have been investigated using multiline techniques in which the polarization signals from many spectral lines are combined in order to enhance extraordinarily the sensitivity of the observations \citep{Semel+Li1996, Donati+etal1997, MartinezGonzalez+etal2008}. The topology of the stellar surface magnetic structure can be then reconstructed using Zeeman-Doppler imaging \citep{Semel1989} from the observation of the rotationally modulated circularly polarized (Stokes V) profiles. Linear polarization (Stokes Q and U) has been 
analyzed for stars hotter than solar-type stars \citep[\eg,][]{Kochukhov+etal2004}, but for cool stars its detection supposes a challenge for state-of-the-art observing facilities. However, it must be included in order to fully characterize the three dimensional structure of the magnetic field, and some pioneering work is now stepping in this direction \citep{Rosen+etal2015}. In addition, linear polarization encodes the information of scattering polarization, which is of key interest to diagnose weak magnetic fields \citep{LopezAriste+etal2011, Ignace+etal2011,MansoSainz+MartinezGonzalez2012}. The detection of the weakly polarized signals from stellar atmospheres requires the development of new instrumentation for the most powerful telescopes. The Mid-resolution InfRAreD Astronomical Spectrograph \citep[MIRADAS,][]{Eikenberry2013} is planned to be installed at the 10.4 m Gran Telescopio Canarias (GTC) in 2019. With its polarimetric capabilities in the near-infrared region and the massive light-gathering power of GTC, MIRADAS will be able to address the study of stellar magnetic fields with an unprecedented detail.   

In this paper we model the spectropolarimetric signals produced by stellar prominences in the \HeI\ 10830 and 5876  \AA\ triplets as they rotate around cool stars. We aim to evaluate the requirements of future spectropolarimeters for detecting and analyzing this phenomenon. In addition, this study provides a benchmark for the interpretation of these future observations. The organization of the paper is as follows: in Section \ref{sect:spectral_synthesis} we describe the code used for the synthesis and the prominence model employed; results are presented in Section \ref{sect:results}; and finally, Section \ref{sect:conclusions} summarizes the conclusions.


\section{Spectral synthesis}
\label{sect:spectral_synthesis}

\subsection{\hazel\ code}
The spectral line synthesis has been performed using the code \hazel\ \citep{AsensioRamos+etal2008}. \hazel\ is able to synthesize
(and invert) Stokes profiles ($I$, $Q$, $U$, $V$) caused by the simultaneous contribution of atomic level polarization and the Hanle and Zeeman effects. The code assumes a model for the triplet system of He \textsc{i} consisting of the first
five terms. The statistical equilibrium equations for the density matrix elements are
solved allowing for quantum interferences among the levels pertaining to each term, which
turns out to be crucial for a correct calculation of the atomic polarization in the He \textsc{i} energy
levels.
Even though the atomic physics is computed in detail, the synthesis is carried out through the 
application of a simple cloud model. All the atoms in a slab of constant physical properties are illuminated by the photospheric continuum coming from below. The emergent polarization signals of the analyzed spectral lines are obtained through the solution of the Stokes vector transfer equation in the slab, assuming that the radiatively induced atomic level polarization is dominated by the photospheric radiation. A solar center-to-limb variation obtained from \citet{Pierce2000} was assumed for the photospheric continuum radiation field. Thus, the presence of sunspots or plages on the stellar surface have been neglected.

\hazel\ has been successfully employed in the study of several solar physics problems, such as determining the magnetic field of spicules \citep{Centeno+etal2010}, analyzing superpenumbral fibrils \citep{Schad+etal2013}, or inferring the magnetic field of solar prominences \citep{OrozcoSuarez+etal2014,MartinezGonzalez+etal2015}. The cloud model is a natural assumption for the study of prominences, since they are clouds of cool plasma located above the stellar surface surrounded by the hot corona. This model is especially good for the interpretation of the \HeI\ 10830 and 5876 (or D$_{\rm 3}$) \AA\ multiplets, since these lines are not generated in the stellar surface and the incident illumination can be assumed as continuum.

\subsection{Star and prominence model}
In this work, we have modeled the intensity and polarimetric signals of the 10830 and 5876 \AA\ multiplets 
(the most important ones in \HeI) produced by stellar prominences as they rotate around their hosting star. The star is characterized by its radius ($R_*$), rotational period ($P$), and angle of inclination ($i$) with respect to the LOS. A cloud representing the prominence is anchored at a given central latitude ($\phi$), while its central longitude ($\xi$) changes according to the rotation of the star. 
At a given time $t$, the central longitude is given by $\xi=2\pi/P(t-t_0)$, where $t_0$ is the time at which the cloud passes across $\xi=0^{\circ}$. In the following we will analyze the full rotation of the prominences. The temporal range studied is $[0,P]$ and we have set $t_0=P/4$. This way, during the first half of the rotation the cloud passes in front of the star. 

The prominence co-rotates with the star at a certain height above the surface ($H$). The size of the cloud is defined by the range in latitude ($\Delta\phi$) and longitude ($\Delta\xi$) that it covers. A homogeneous magnetic field with strength $B$ is assumed for all the cells of the prominence. The orientation of the magnetic field is described by its inclination with respect to the local vertical ($\theta_B$), where $\theta_B=0^{\circ}$ corresponds to a radially outwards direction (from the center of the star), and azimuth $\chi_B$ defined in the perpendicular plane. We assume that the dominant broadening is produced by a thermal Doppler velocity of 6.4 km s$^{-1}$ for helium,
which is equivalent to a prominence temperature of 10,000 K. The optical depth in the \HeI\ 10830 and 5876 \AA\ multiplets was set to one. The cloud properties are constant in time. 

The prominence was sampled by dividing its area into a grid with $n_{\phi}$ positions in latitude and $n_{\xi}$ positions in longitude, making a total of $n_{\phi}\times n_{\xi}$ cells. For every temporal step, the synthesis of the spectral lines was performed at each one of these cells. The integrated prominence spectrum was then calculated by multiplying the emergent Stokes profiles from each cell by its corresponding area in the plane of the sky (computed from its distance to the neighbor cells) and adding it to the contributions from other cells. On the other hand, the radiation emerging from the stellar disk is simply the continuum intensity (unity) for all wavelengths and a zero value for Stokes $Q$, $U$, and $V$, since we are assuming that there are not active regions on the stellar surface. The integrated stellar spectrum (including the prominence) is given by the sum of the integrated prominence spectrum and the continuum integrated for the stellar disk, excluding the non-visible regions (from the star or the prominence) due to eclipses. Finally, the intensity spectrum is normalized to unity in the continuum and the Stokes parameters $Q$, $U$, and $V$ are multiplied by the same normalizing factor.

\subsection{Geometry of the problem}

Characterizing the scattering event produced at the slab requires the definition of several angles which describe the location of the slab with respect to the star and the LOS, and the orientation of the magnetic field. Figure 3 from \citet{AsensioRamos+etal2008} illustrates the angles that need to be provided to \hazel\ code in order to compute the synthesis of the spectral lines.

We start considering one of the cells of the prominence with latitude $\phi_j$ and longitude $\xi_j$. This latitude is the angle between the star equator and the position of the cell, while the longitude is the angle measured along the equator between the cell and the center of the stellar disk. If the angle of inclination $i$ is different from $90^{\circ}$, the cell position will present a different apparent latitude $\phi_j^{0}$ and longitude $\xi_j^{0}$ on the plane of the sky. These coordinates are defined with respect to the apparent equator, located at the middle of the visible stellar disk. They are given by 

\begin{equation}
\begin{aligned} 
\label{eq:LOS}
&\sin\phi_j^{0}=\sin\phi_j\cos (90^{\circ}-i)-\cos\phi_j\sin (90^{\circ}-i)\sin (\xi_j+90^{\circ}),\\
&\cos(\xi_j^{0}+90^{\circ})=\cos\phi_j\cos (\xi_j+90^{\circ})/\cos\phi_j^{0},\\
\end{aligned}
\end{equation}

\noindent where the angles are introduced in degrees. Note that we have assumed that the axis of the star is contained in the plane defined by the LOS and the vertical direction. We have neglected any change in its inclination on the plane of the sky.  

In order to simplify the calculations, it is advantageous to define three 
different reference frames. The ``local reference frame'' moves with the slab and it is described by the unity vectors $(\hat{x}_{\rm sl},\hat{y}_{\rm sl},\hat{z}_{\rm sl})_{\rm sl}$, where the subindex ``sl'' clarifies that this reference frame is
fixed with the slab. The vector $\hat{z}_{\rm sl}$ is directed radially outwards from the center of the star, $\hat{y}_{\rm sl}$ points towards the north pole, and $\hat{x}_{\rm sl}$ is the orthonormal right-handed vector. The ``inclined local reference frame'' $(\hat{x}_{\rm sl}^i,\hat{y}_{\rm sl}^i,\hat{z}_{\rm sl})_{\rm sl}^i$ is defined after the inclination of the stellar axis has been taken into account. Thus, the vector $\hat{y}_{\rm sl}^i$ is not anymore directed to the north pole of the star. Instead, it points to the apparent north pole on the plane of the sky. The unity vector $\hat{z}_{\rm sl}^i$ is also directed radially outwards from the center of the star and $\hat{x}_{\rm sl}^i$ is the orthonormal right-handed vector. Finally, the stationary reference frame $(\hat{x}_{0},\hat{y}_{0},\hat{z}_{0})_0$ is placed at the center of the star, with $\hat{x}_{0}$ pointing in the direction of the LOS, $\hat{z}_{0}$ is the vertical, and $\hat{y}_{0}$ points to the west. The transformation matrix between the stationary reference frame and the inclined local reference frame is


\begin{equation}
\label{eq:LOS_to_inclined}
D^i = 
\begin{bmatrix} -\sin\xi_j^{0} & \cos\xi_j^{0} & 0 \\ -\sin\phi_j^{0}\cos\xi_j^{0} & -\sin\phi_j^{0}\sin\xi_j^{0} & \cos\phi_j^{0} \\ \cos\phi_j^{0}\cos\xi_j^{0} & \cos\phi_j^{0}\sin\xi_j^{0} & \sin\phi_j^{0} \end{bmatrix} 
.
\end{equation}

\noindent Similarly, the transformation matrix between the stationary reference frame and the local reference frame $(D)$ is retrieved after substituting $\phi_j^{0}$ and $\xi_j^{0}$ by $\phi_j$ and $\xi_j$, respectively. According to Figure 3 from \citet{AsensioRamos+etal2008}, the LOS is characterized by the angles $\theta$ and $\chi$, where $\theta$ is the angle between the local vertical and the LOS and $\chi$ is the angle between the $\hat{x}_{sl}^i$ direction and the plane formed by the LOS and the local vertical. Using matrix $D^i$, it is straightforward to obtain the LOS vector $(1,0,0)_0$ in the inclined local reference frame and retrieve the angles $\theta$ and $\chi$ as  

\begin{equation}
\begin{aligned} 
\label{eq:theta_chi}
&\cos\theta=\cos\xi_j^{0}\sin(90^{\circ}-\phi_j^{0}),\\
&\tan\chi=\sin\phi_j^{0}\cos\xi_j^{0}/\sin\xi_j^{0}.\\
\end{aligned}
\end{equation}

\begin{figure}[!t] 
 \centering
 \includegraphics[width=9cm]{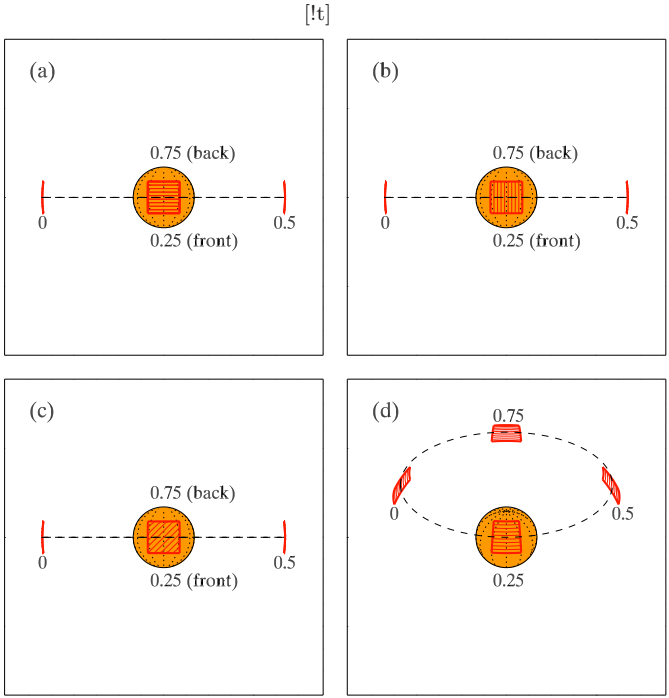}
  \caption{Representation of the star (yellow sphere) and prominence (red lines) model for four configurations. Panel $a$: $i=90^{\circ}$, $\phi=0^{\circ}$,  $H=3R_*$, $\chi_B=0^{\circ}$; panel $b$: same as $a$ but $\chi_B=90^{\circ}$, panel $c$: same as $a$ but $\chi_B=45^{\circ}$; panel $d$: $i=60^{\circ}$, $\phi=30^{\circ}$,  $H=3R_*$, $\chi_B=0^{\circ}$. The prominence is illustrated at four different positions corresponding to rotational phases $0$, $0.25$, $0.5$, and $0.75$ (the rotational phase is printed next to the prominence). The thin red lines inside the prominences are parallel to the direction of their magnetic field. The black dashed lines show the trajectory of the prominence.}
  
  \label{fig:modelo}
\end{figure}

In addition, the (arbitrary) positive reference direction for Stokes $Q$ is defined by the direction of the vector $\mathbf{e}_1$ in Fig. 3 from \citet{AsensioRamos+etal2008}, which is measured from the plane formed by the vertical and the LOS. 
In our case, we have chosen to set $\mathbf{e}_1$ in the direction $\hat{y}_{0}$. It is obtained as

\begin{equation}
\tan\gamma=\tan (90^{\circ}-\phi_j^{0})\sin\xi_j^{0}.
\label{eq:gamma}
\end{equation}

The synthesized profiles will also be affected by the velocity on the LOS. We have assumed that the prominence is co-rotating with the star with no proper motion. This way, the only contribution to the LOS velocity is that produced by the rotation. In the local reference frame it is given by
\begin{equation}
\mathbf{v}_{\rm rot}=\frac{2\pi}{P}(R_*+H)\cos\phi_j\hat{x}_{\rm sl}.
\label{eq:v_rot}
\end{equation}

\noindent The rotational velocity was rotated an angle $90^{\circ}-i$ around the $y_0$ axis $(0,1,0)_0$ in order to account for the inclination of the star axis with respect to the LOS, and the resulting velocity was then projected on the LOS $(1,0,0)_0$ using the transpose of matrix $D$.

The last geometrical consideration refers to the magnetic field orientation. It is described by the angles $\theta_B$ and $\chi_B$, which are defined in the local reference frame. $\theta_B$ is the angle between $\hat{z}_{\rm sl}$ and the magnetic field vector, while $\chi_B$ indicates the angle between the horizontal magnetic field and the direction $\hat{x}_{\rm sl}$. The components of the magnetic field in the local reference frame have been rotated according to the inclination $i$ and converted to the inclined local reference frame. Then, the angles $\theta_B^i$ and $\chi_B^i$ were computed using similar definitions to the corresponding angles in the local reference frame. Note that in both systems of reference the $z$ axis points radially outwards, so $\theta_B^i=\theta_B$. On the other hand, the horizontal components of the magnetic field will be generally different in both reference frames and the angle $\chi_B^i$ will depend on $\chi_B$, $i$, $\theta_j$, and $\xi_j$. The angles $\theta_B^i$ and $\chi_B^i$ need to be provided to \hazel.

\begin{figure*}
 \centering
 \includegraphics[width=17cm]{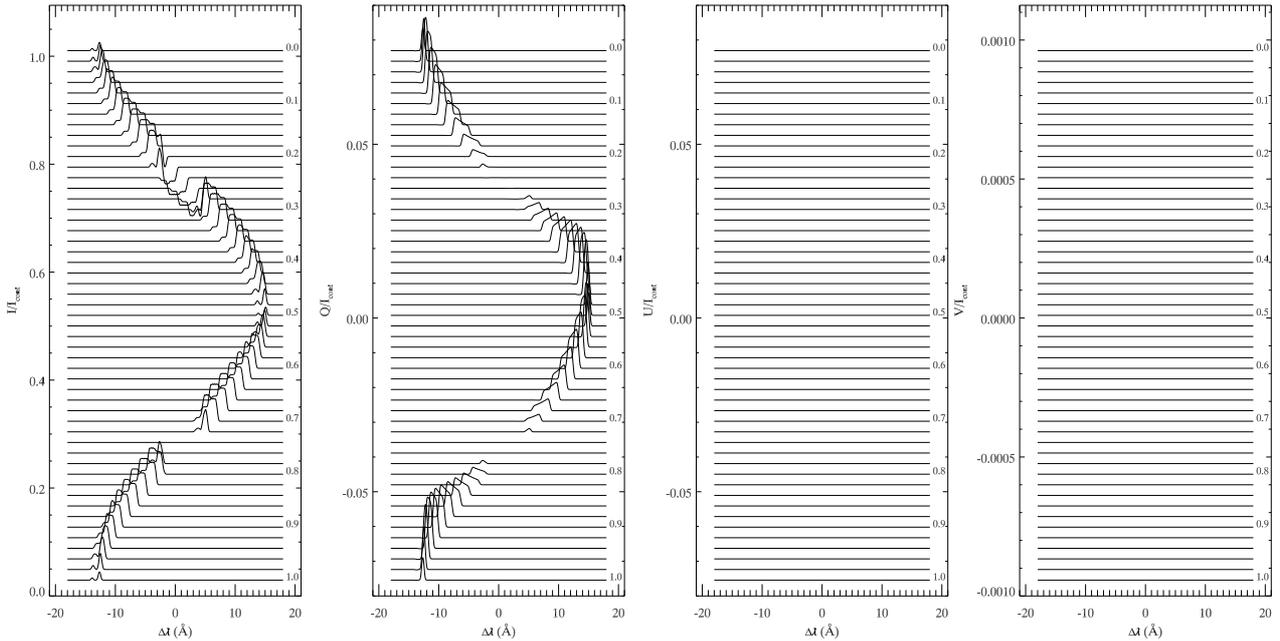}
  \caption{Synthetic Stokes parameters in the \HeI\ 10830 \AA\ triplet for a star with $i=90^{\circ}$ and a prominence rotating in the equatorial plane ($\phi=0^{\circ}$) with $\Delta\phi=12^{\circ}$, $\Delta\xi=12^{\circ}$, $B=0$ G, and $H=3R_*$. Successive profiles are shifted downwards for graphics purposes, with the rotational phase (noted right to some profiles) increasing from top to bottom.}
  \label{fig:stokes_prom1}
\end{figure*}

\section{Results}
\label{sect:results}

In this work we have evaluated the variation of the Stokes parameters from the \HeI\ 10830 and 5876 \AA\ triplets during one full rotation of a prominence around a star. Since state-of-the-art instruments are still not able to measure these signals, this analysis has been designed following an exploratory approach. We have carried out a parametric study of the synthetic \HeI\ 10830 \AA\ line generated under different configurations for the star-prominence system. Several parameters have been sampled, including the inclination of the star and the latitude and magnetic field of the prominence. In all cases the values of each parameter have been chosen to lie within reasonable values according to our current understanding of the physical properties of stellar prominences.\footnote{To explore the space
of parameters, we offer the code used to generate the synthetic signals in \texttt{http://github.com/tfelipe/stellar\_prominence}, which makes use of \hazel, whose
latest version can be found in \texttt{http://github.com/aasensio/hazel}.}

Figure \ref{fig:modelo} illustrates the prominence models explored. In all panels the prominence is plotted at four different positions, corresponding to rotational phases of $0$, $0.25$, $0.5$, and $0.75$. The thin red lines inside the prominences are parallel to the direction of their magnetic field. Prominences are expected to be trapped in long loops surrounded by the corona. At the top of the loop, we can consider that the magnetic field is mostly horizontal and, thus, in all cases we have selected an horizontal magnetic field ($\theta_B=90^{\circ}$). The estimation of the prominence field strength is intricate, since no direct measurements have been obtained so far. In order to hold the prominence, the magnetic tension at the top of the coronal loop must be high enough to outweigh the local effective gravity. This can be expressed as

\begin{equation}
\label{eq:Bprom}
\frac{B_{\rm min}^2}{\mu_0r_{\rm c}}=\rho g_{\rm eff}, 
\end{equation}

\noindent where $B_{\rm min}$ is the minimum value of the magnetic field strength that can sustain the prominence, $\mu_0$ is the magnetic permeability, $r_{\rm c}$ is the radius of the curvature at the top of the loop, $\rho$ is the mass density, and $g_{\rm eff}$ is the effective gravity, which includes the contribution of the centrifugal acceleration. For a prominence located at the equatorial plane the effective gravity is purely radial and can be obtained as

\begin{equation}
\label{eq:geff}
g_{\rm eff}=-\frac{GM_*}{R_*+H}+\omega^2(R_*+H),
\end{equation}

\noindent where $G$ is the gravitational constant, $M_*$ is the stellar mass, and $\omega=2\pi/P$ is the rotational angular velocity. When the prominence is located at the co-rotation radius, gravity is balanced by centrifugal acceleration and a negligible field strength is necessary to hold the prominence. On the other hand, the larger the radial distance of the prominence, the higher the magnetic tension which is needed to compensate the effective gravity. We have estimated the minimum field strength expected for a stellar prominence located at the equatorial plane in a star with the properties of AB Dor. We set $R_*=0.98R_{\odot}$ \citep{CollierCameron+Foing1997, Maggio+etal2000}, $M_*=0.86M_{\odot}$ \citep{Guirado+etal2006}, $P=0.514$ days \citep{Innis+etal1988}. For the distance of the cloud to the surface of the star we have selected $H=3R_*$, which lies within the radial range where prominences have been detected in AB Dor \citep[\eg,][]{Donati+etal1999}. Following \citet{Donati+etal2000}, we have taken $r_{\rm c}=0.3R_*$ and $\rho =5\times10^{-14}$ g cm$^{-3}$. The minimum magnetic field strength required for holding this prominence, given by Equation \ref{eq:Bprom}, is slightly higher than 7 G. Therefore, a 10 G field strength has been chosen for the magnetized prominence cases. The longitudinal extent of the prominence has been chosen in such a way that when the cloud is located at the center of the stellar disk it covers 25\% of its area \citep[\eg][]{CollierCameron+Robinson1989a}.

\begin{figure*}
 \centering
 \includegraphics[width=17cm]{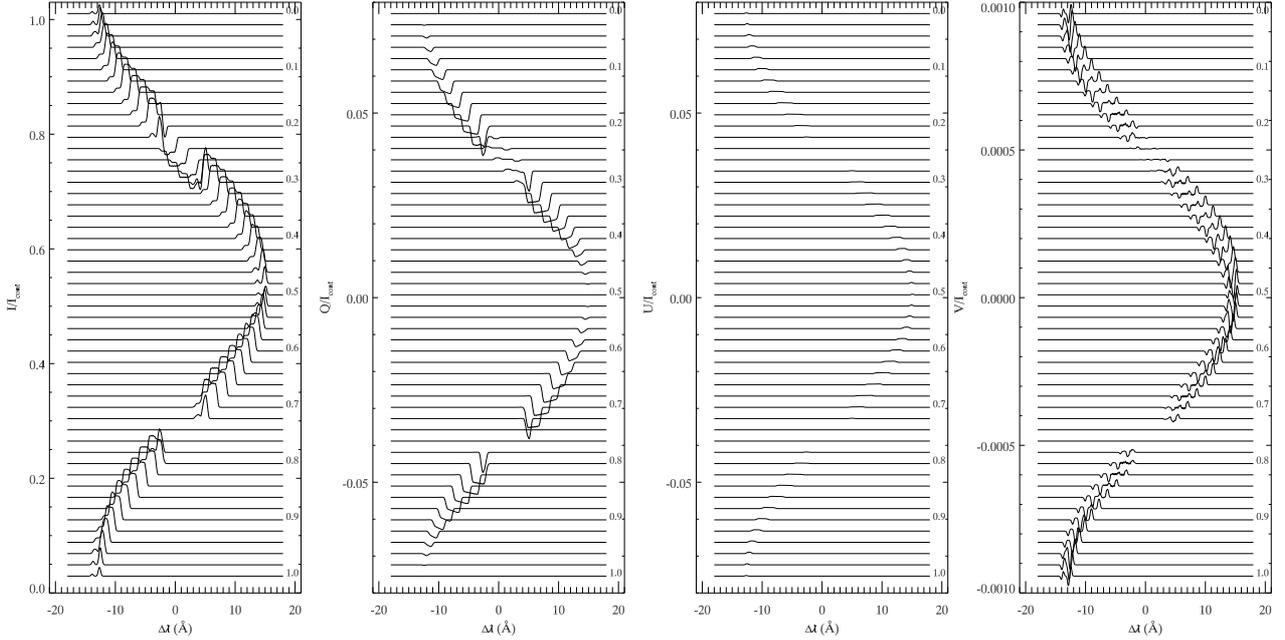}
  \caption{Same as Figure \ref{fig:stokes_prom1} but for a prominence magnetic field given by $B=10$ G, $\theta_B=90^{\circ}$,  $\chi_B=0^{\circ}$. }
  \label{fig:stokes_prom2}
\end{figure*}

\begin{figure*}
 \centering
 \includegraphics[width=17cm]{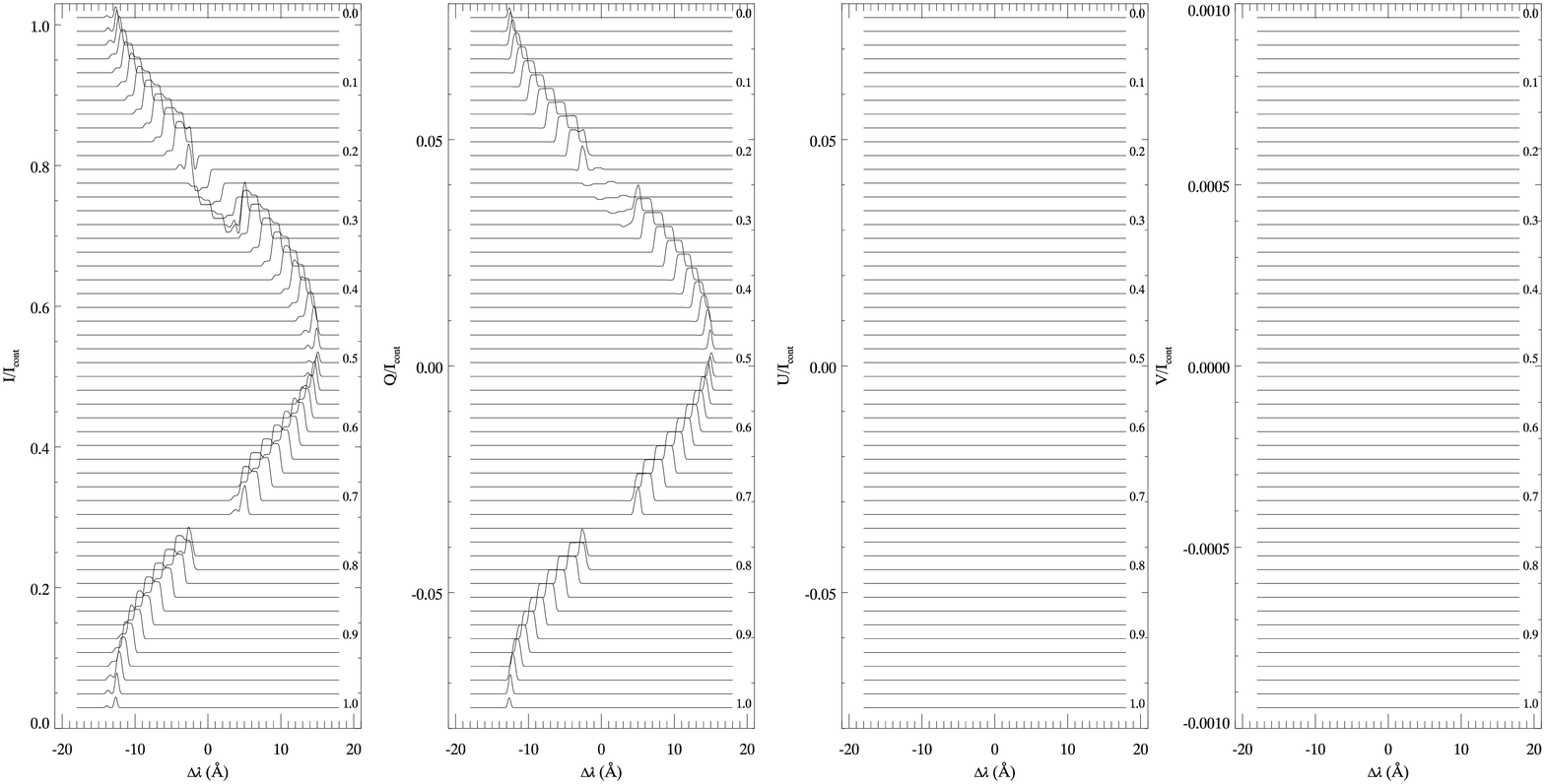}
  \caption{Same as Figure \ref{fig:stokes_prom1} but for a prominence magnetic field given by $B=10$ G, $\theta_B=90^{\circ}$,  $\chi_B=90^{\circ}$. }
  \label{fig:stokes_prom3}
\end{figure*}

\begin{figure*}
 \centering
 \includegraphics[width=17cm]{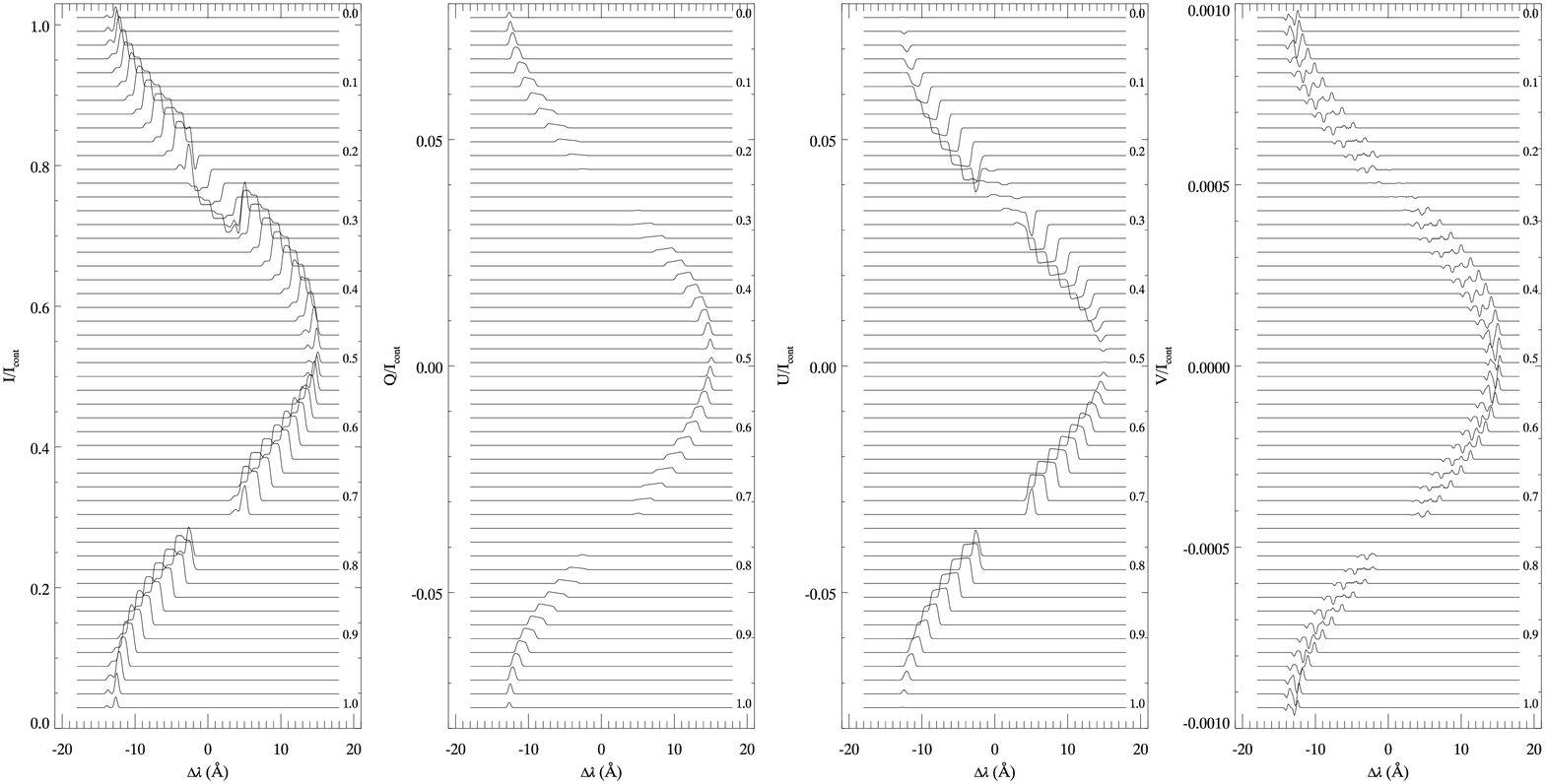}
  \caption{Same as Figure \ref{fig:stokes_prom1} but for a prominence magnetic field given by $B=10$ G, $\theta_B=90^{\circ}$,  $\chi_B=45^{\circ}$. }
  \label{fig:stokes_prom4}
\end{figure*}

\subsection{Time evolution of the Stokes parameters}

\subsubsection{Field free prominence}

Figure \ref{fig:stokes_prom1} shows the evolution of the Stokes profiles from the \HeI\ 10830 \AA\ triplet generated by a prominence that rotates around a star with its rotation axis perpendicular to the LOS ($i=90^{\circ}$). Stokes parameters have been integrated for all the system and the continuum intensity has been normalized to one. The prominence is located at a latitude $\phi=0^{\circ}$. It rotates in the equatorial plane, crossing the stellar disk when it passes in front of the star and hiding completely behind the star for rotational phases between 0.73 and 0.77 (see panel $a$ from Figure \ref{fig:modelo}). At phase 0.0 the prominence is located out of the stellar disk and it produces an emission feature due to the photons that are scattered into the LOS by the cloud. Its wavelength displacement is given by the rotation velocity along the LOS. The emission feature shows two peaks, the red one presenting a higher intensity, produced by the different components of the \HeI\ 10830 \AA\ triplet. It drifts across the profile and when the cloud starts to eclipse the star an absorption feature appears in the red side of the detected intensity signature. For rotational phases between 0.19 and 0.22 the line shows simultaneously absorption and emission, while a pure absorption transient drifts across the profile between phases 0.23 and 0.27. During this time all the surface of the cloud is blocking the light from the star. The prominence scatters many of the photons coming from the surface out of the LOS, reducing the intensity. During the rest of the rotation of the prominence around the star, an emission feature drifts across the profile according to the changes in the rotational velocity along the LOS, except for the time that the cloud is occulted by the star. For these rotational phases (between 0.73 and 0.77) there is no trace of the prominence in the Stokes profiles.
   
\begin{figure*}
 \centering
 \includegraphics[width=17cm]{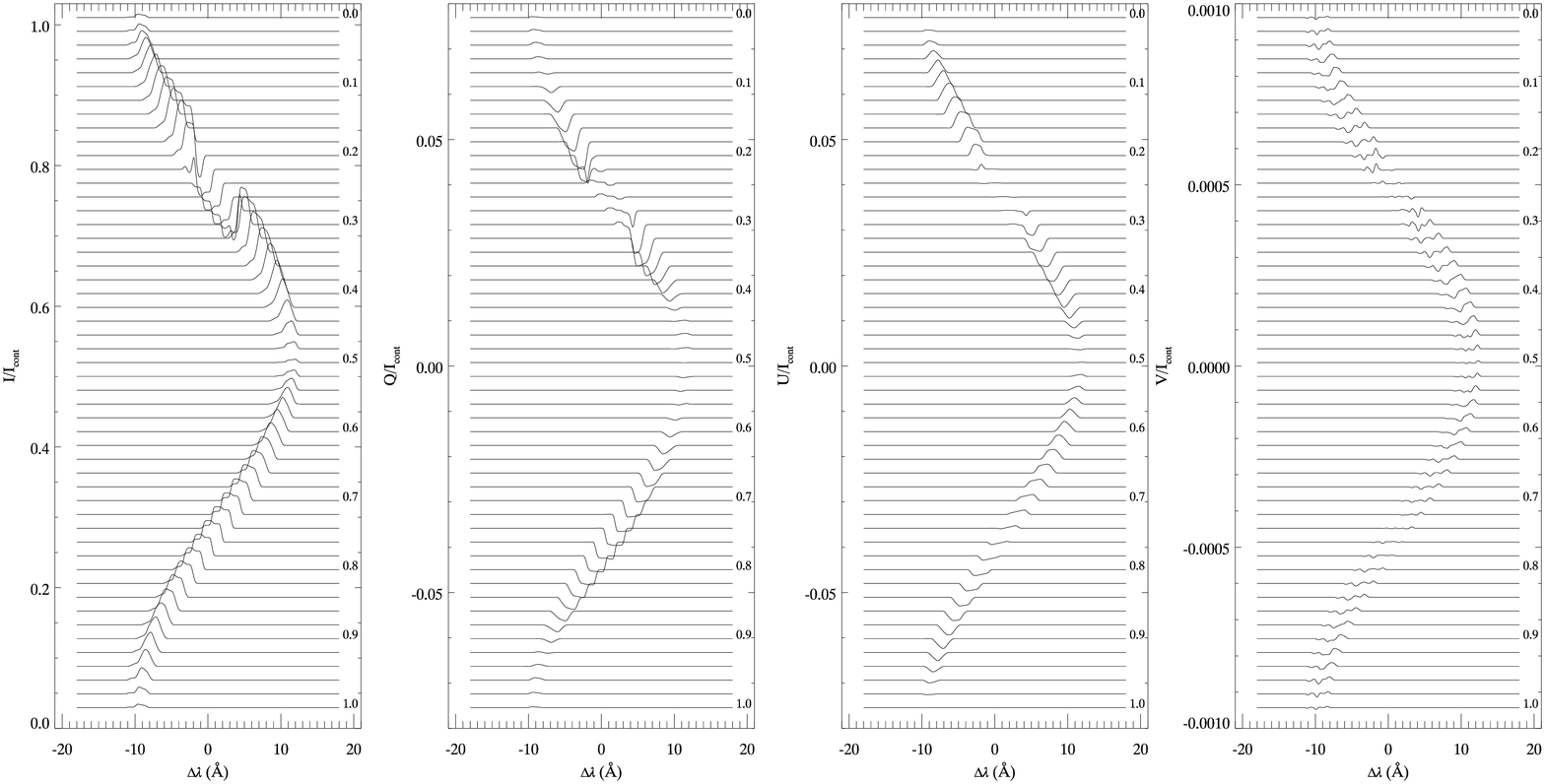}
  \caption{Same as Figure \ref{fig:stokes_prom1} but for a star with $i=60^{\circ}$ and a prominence at $\phi=30^{\circ}$ with $\Delta\phi=13^{\circ}$, $\Delta\xi=13^{\circ}$, $B=10$ G, $\theta_B=90^{\circ}$,  $\chi_B=0^{\circ}$ and $H=3R_*$.}
  \label{fig:stokes_prom5}
\end{figure*}

Although we are assuming a non-magnetic prominence, the scattering of light coming from the stellar photosphere in the cloud plasma can produce linear polarization in the \HeI\ 10830 \AA\ triplet (and other spectral lines and continuum). This linear polarization appears in Stokes $Q$ parameter and it is maximum when the LOS forms a 90$^{\circ}$ angle with the direction the cloud is illuminated (local vertical). In this case its amplitude is about 1.6\% of the intensity continuum. The contribution of the scattering polarization to the signal will be stronger when the prominence is out of the stellar disk, for rotational phases around 0.0 (1.0) and 0.5. The distribution of the linearly polarized signal between Stokes $Q$ and $U$ depends on the geometry of the scattering. According to our selection of the reference direction for Stokes $Q$ (positive for polarization along $\hat{y}_{0}$), the prominence located at the equator only yields positive values of Stokes $Q$, while Stokes $U$ vanishes. A scattering event with the LOS in the same direction of the local vertical produces no scattering polarization due to its symmetry. This way, polarization signals are negligible when the cloud is in front of the star (rotational phases between 0.23 and 0.27).

\begin{figure*} 
 \centering
 \includegraphics[width=17cm]{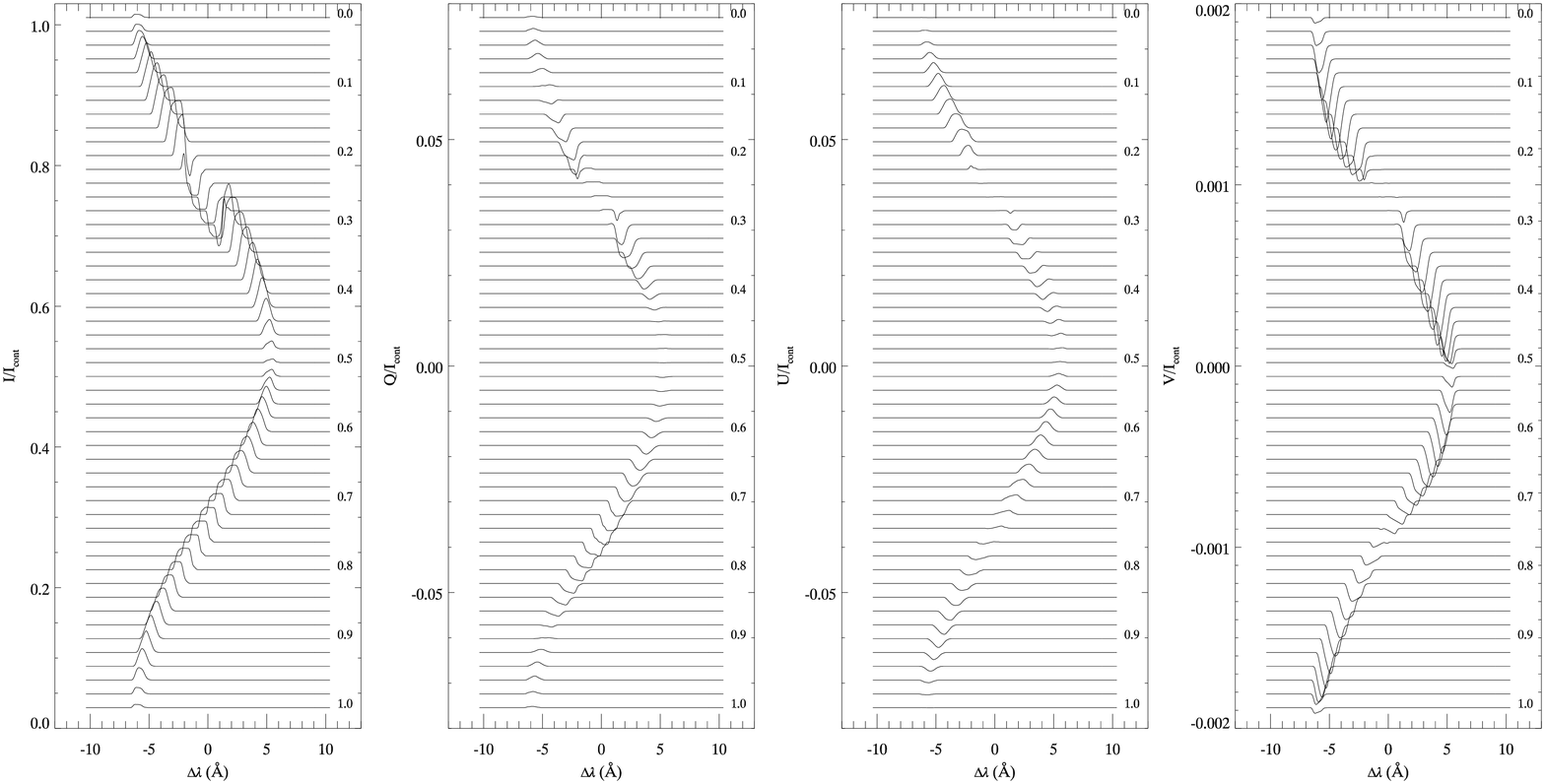}
  \caption{Same as Figure \ref{fig:stokes_prom5} but for \HeI\ 5876 \AA.}
  \label{fig:stokes_prom6}
\end{figure*}

\subsubsection{Magnetized prominences}

Figures \ref{fig:stokes_prom2}, \ref{fig:stokes_prom3}, and \ref{fig:stokes_prom4} show three cases with the same geometry of the prominence discussed in the previous section but with a magnetic field strength of 10 G. They differ in the orientation of the prominence magnetic field. In all cases it is horizontal, but their azimuths $\chi_B$ are $0^{\circ}$ (Figure \ref{fig:stokes_prom2}), $90^{\circ}$ (Figure \ref{fig:stokes_prom3}), and $45^{\circ}$ (Figure \ref{fig:stokes_prom4}). Their configurations are illustrated in panels $a$, $b$, and $c$ from Figure \ref{fig:modelo}, respectively. In the three models the evolution of the intensity profile shows no significant changes with respect to the unmagnetized prominence. However, the presence of a weak magnetic field modifies the scattering process and leads to a different linear polarization pattern through the so-called Hanle effect. The Hanle effect tends to reduce the linear polarization and produces a rotation of the plane of polarization, and in these cases the Stokes $Q$ profiles present lower amplitudes than in the unmagnetized prominence.

When the magnetic field is directed along parallels with constant latitude ($\chi_B=0^{\circ}$, Figure \ref{fig:stokes_prom2}), the Stokes $Q$ profiles only show negative values. Their maximum amplitude (around 0.6\% of the intensity continuum) is obtained when the prominence is near the stellar disk. They are greatly reduced when the cloud eclipses the star but, contrary to the field free case, they are not negligible. The presence of a magnetic field perpendicular to the LOS breaks the symmetry of the scattering event produced when the cloud is at the center of the stellar disk, generating some linear polarization in the forward scattering process. The Stokes $U$ and $V$ profiles show very low signals. In the case of Stokes $V$, they are produced by the Zeeman effect. Circular polarization (Stokes $V$) is generated through Zeeman effect when some component of the magnetic field is directed along the LOS. For the prominence with $\chi_B=0^{\circ}$, the LOS magnetic field is different from zero when the cloud is out of the center of the stellar disk. The maximum Stokes $V$ signals are found for rotational phases $0$ ($1$) and $0.5$, but even in those cases their magnitude are very low (note the smaller scale used for the plot of Stokes $V$). 

For a prominence magnetic field with $\chi_B=90^{\circ}$ (Figure \ref{fig:stokes_prom3}) the linear polarization signal produced by the Hanle effect is also shown in the Stokes $Q$ parameter. The scattering process at the cloud is similar to that discussed in the previous paragraphs, but in this case the orientation of the magnetic field produces positive values of Stokes $Q$. The magnetic field is always almost perpendicular to the LOS and, thus, the circular polarization is negligible for all rotational phases.

Figure \ref{fig:stokes_prom4} illustrates the prominence with its magnetic field oriented in the direction given by $\chi_B=45^{\circ}$. In this case, most of the scattering polarization appears in the Stokes $U$ parameter. For the rotational phases where the cloud is in front of the star (between 0 and 0.5) its signal is negative (except when prominence is on the stellar disk, where some small positive signals are found). On the other hand, Stokes $U$ parameter shows positive values when the cloud is behind the star (rotational phases between 0.5 and 1). Stokes $Q$ profiles show some positive signals. With regards to the circular polarization, a small signal is obtained in Stokes $V$. It is smaller than that found for the prominence with $\chi_B=0^{\circ}$, since in this case the magnetic field component along the LOS is lower.

For a comprehensive view of the problem, we have also studied a more realistic case where the inclination of the star and latitude of the prominence were chosen in order to mimic AB Dor. The inclination was set to 60$^{\circ}$ and the latitude to $\phi=30^{\circ}$, so that the cloud crosses the center of the stellar disk. The period and height of the prominence are equal to those selected for the previous cases, and the magnetic field is also horizontal (in the local reference frame), with 10 G strength and $\chi_B=0^{\circ}$. This configuration corresponds to panel $d$ from Figure \ref{fig:modelo}, and the evolution of the Stokes profiles is shown in Figure \ref{fig:stokes_prom5}.

The intensity profiles show a strong resemblance with those presented for the prominences located at the equatorial plane. An emission feature appears at the blue side of the line and drifts across the profile. During the rotational phases when  the cloud covers part of the stellar disk an absorption feature appears in the intensity profiles. According to the inclination of the rotation axis of the star and the latitude of the prominence, the latter is never occulted behind the stellar disk (see Figure \ref{fig:modelo}). The emission feature is visible drifting from the red to the blue for all rotational phases between 0.5 and 1, as opposed to the previous cases, where it disappears for rotational phases around 0.75. The maximum magnitude of the Doppler shift (approximately 10 \AA\ at rotational phases 0 and 0.5) is lower than in the cases where the prominence is located at the equatorial plane (approximately 13 \AA). A cloud located at the same distance from the center of the star ($R_*+H$) but at a certain latitude is closer to the rotational axis and its rotational velocity is lower. In addition, the rotational velocity is never directed along the LOS.

The linear polarization produced by the scattering is distributed between the Stokes $Q$ and $U$ parameters. Stokes $Q$ is mainly negative, small positive signals are only found when the prominence is eclipsing the star and around rotational phases 0.0 and 0.5. The magnitude of Stokes $U$ is slightly lower than that of Stokes $Q$ and its sign switches between positive and negative as the cloud completes its rotation around the star. Stokes $V$ also shows a very low signal.

The \HeI\ 5876 \AA\ triplet has also been synthesized for the latter case (a star with $i=60^{\circ}$ and prominence at $\phi=30^{\circ}$, illustrated in panel $d$ from Figure \ref{fig:modelo}). Figure \ref{fig:stokes_prom6} shows the results. The Stokes profiles are qualitatively similar to those obtained for the \HeI\ 10830 \AA\ triplet, although some differences arise. First, the Doppler shift of this line is lower since its wavelength is shorter. Second, the amplitude of the linearly polarized signals (Stokes $U$ and $Q$) are lower. Third, Stokes $V$ shows a symmetric profile (instead of the antisymmetric profiles found for \HeI\ 10830 \AA) since this multiplet is dominated by the alignment-to-orientation transfer mechanism \citep{Kemp+etal1984}. Stokes $V$ amplitude is higher than that of the \HeI\ 10830 \AA\ triplet, but still lower than the signals estimated for the linear polarization.

\begin{figure*}
 \centering
 \includegraphics[width=17cm]{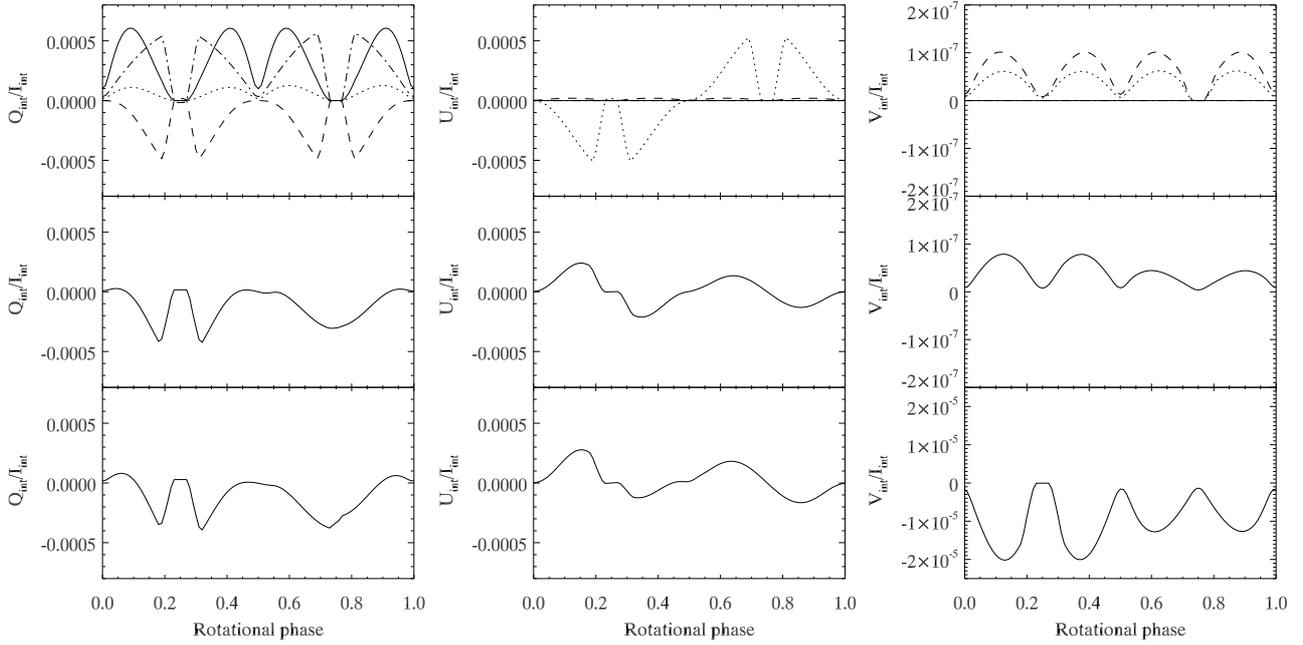}
  \caption{Wavelength integrated polarization. Top panels: \HeI\ 10830 \AA\ signals for a prominence located at the equatorial plane with $B=0$ G (solid line), $B=10$ G and $\chi_B=0^{\circ}$ (dashed lines), $B=10$ G and $\chi_B=45^{\circ}$ (dotted lines), and $B=10$ G and $\chi_B=90^{\circ}$ (dash-dotted lines); middle panels: \HeI\ 10830 \AA\ signals for a star with $i=60^{\circ}$ and a prominence at $\phi=30^{\circ}$, $B=10$ G,  $\chi_B=0^{\circ}$; bottom panels: \HeI\ 5876 \AA\ signals for the same prominence from middle panels. }
  \label{fig:polarizacion_integrada}
\end{figure*}

\subsubsection{Integrated polarization}
\label{sect:integrated}
Figure \ref{fig:polarizacion_integrada} shows the temporal variation of the wavelength integrated polarization during a full rotation of the prominence for all the cases discussed in the previous sections. The integration is performed on the wavelength range $[\lambda_0-\Delta\lambda, \lambda_0+\Delta\lambda]$, where $\lambda_0$ in the center of the spectral line and $\Delta\lambda$ gives the width of the integration. It has been chosen to includes all the stellar rotation profile. For $\lambda_0=10830$  \AA\ (top and middle panels) we set $\Delta\lambda=16$ \AA, while in the case of $\lambda_0=5876$ \AA\ (bottom panels) the wavelength range is $\Delta\lambda=7$ \AA.

The top panel of Figure \ref{fig:polarizacion_integrada} illustrate the four cases with the prominence located on the rotational plane. In the field free case (solid line) the polarization signal is found in Stokes $Q$, as previously seen in Figure \ref{fig:stokes_prom1}. The presence of a magnetic field dramatically changes the wavelength integrated profiles. The maximum polarization is obtained at the rotational phases when the prominence is near the stellar disk, but not covering it. The field modifies the geometry of the scattering. Depending on the orientation of the magnetic field the distribution of the linear polarization signals between Stokes $Q$ and $U$ differs, although the total linear polarization is similar. For example, the signal in the cases with $\chi_B=0^{\circ}$ and $\chi_B=90^{\circ}$ appears mainly in Stokes $Q$, but they have opposite amplitude. The prominence with $\chi_B=45^{\circ}$ shows a similar pattern in Stokes $U$. 

As expected, the circular polarization signal is very low in the cases where there is a longitudinal component of the magnetic field (with respect to the LOS) and it vanishes when it is zero. 

The rotation of the prominence hosted by a star with inclination $i=60^{\circ}$ exhibit a similar profile in \HeI\ 5876 and 10830 \AA\ triplets. The wavelength integrated Stokes $Q$ during the first half of the rotation is also similar to the equatorial prominence with $\chi_B=0^{\circ}$, since in those cases we have also impose this magnetic field orientation. However, a non-negligible value is obtained in Stokes $U$. In the second half of the rotation the prominence is never occulted behind the star, and the variation of the wavelength integrated linearly polarized signature is more smooth.


\subsubsection{MIRADAS spectral resolution}
\begin{figure*}
 \centering
 \includegraphics[width=17cm]{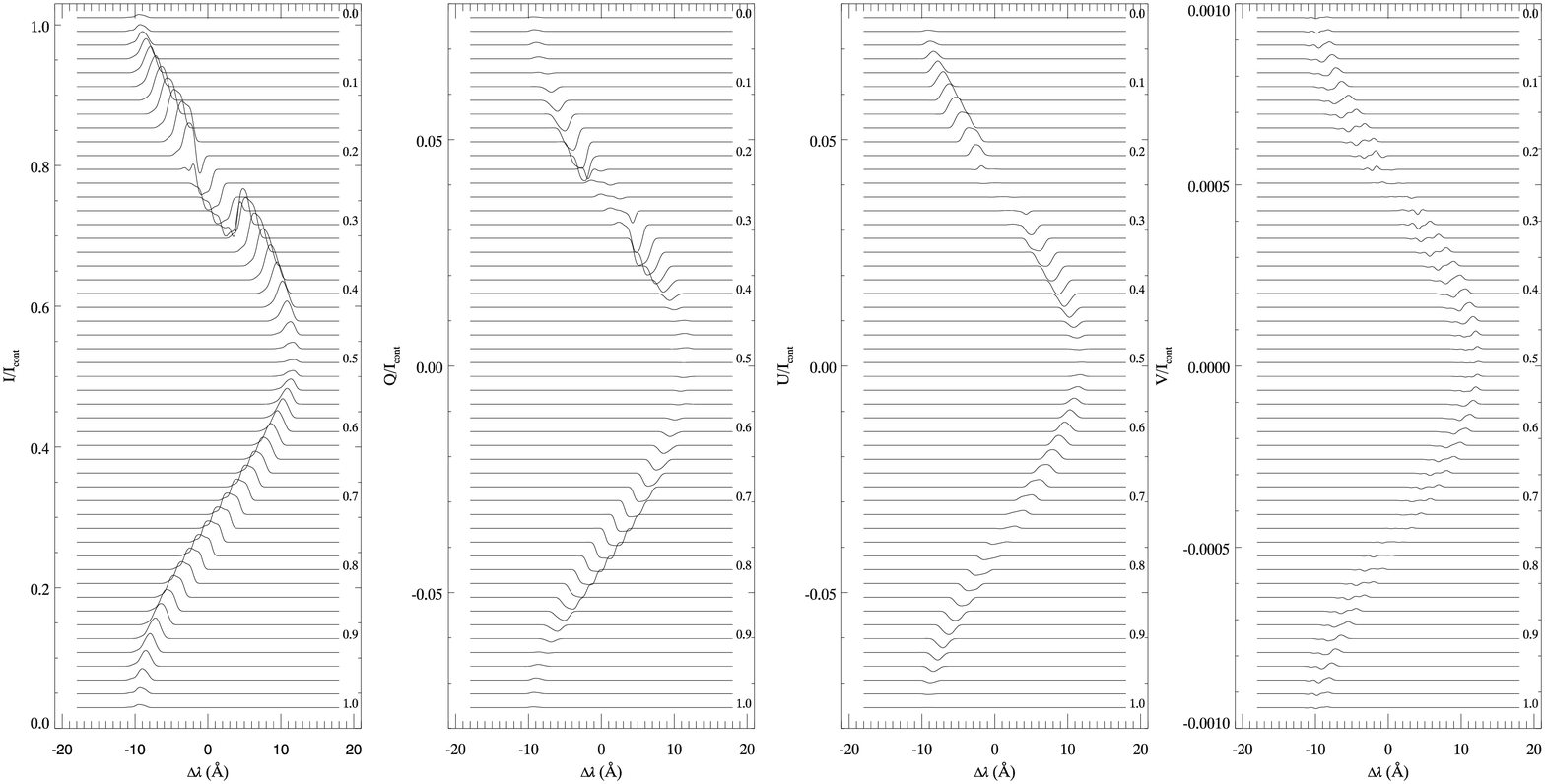}
  \caption{Same as Figure \ref{fig:stokes_prom5} but with the spectral resolution degraded to R=20,000.}
  \label{fig:stokes_prom5_MIRADAS}
\end{figure*}

All the previous calculations have infinite spectral resolution, and
the signals are expected to decrease when observed with a finite resolution
spectropolarimeter. In order to permit a direct comparison with MIRADAS, the spectral resolution has been degraded to R=20,000,
its nominal value. The profiles of the four Stokes parameters have been convolved with a Gaussian of FWHM given by the resolution at the corresponding wavelength. Figure \ref{fig:stokes_prom5_MIRADAS} shows the degraded Stokes parameters for the case previously illustrated in Figure \ref{fig:stokes_prom5}. Thanks to the high spectral resolution of MIRADAS, the profiles barely change after spectral degradation. The variation of the amplitude of the intensity is negligible and the change in the amplitude of Stokes $Q$ and $U$ parameters is about 2\%. Even the most fainter features of the undegraded profiles, such as the low Stokes $Q$ signal when the prominence covers the stellar disk or the coexistence of low intensity emission and absorption when only a section of the prominence is in front of the stellar disk, are still clearly visible in Figure \ref{fig:stokes_prom5_MIRADAS}. Stokes $V$ profiles are especially affected by the degraded resolution, since the circularly-polarized signature is more complex than those seen in Stokes $Q$ and $U$. Some of the sign reversals of Stokes $V$ which can be seen in Figure \ref{fig:stokes_prom5} are lost in the degraded profiles.

\section{Discussion and conclusions}
\label{sect:conclusions}

The full characterization of stellar magnetic fields requires the analysis of the four Stokes parameters. However, due to the limited sensitivity of current infrastructures, studies to date have been restricted to the observation of only Stokes $I$ and $V$ parameters. The interpretation of the full polarimetric information can potentially open a window for probing magnetic fields even in stellar coronae. This process will greatly benefit from the analysis of polarizing phenomena that are well-known for solar physics studies, such as the scattering polarization and the Hanle effect. Some authors have already proposed the possibility of applying the Hanle effect to diagnose stellar magnetism \citep{Ignace+etal1997,Ignace+etal1999, Nordsieck2001, MansoSainz+MartinezGonzalez2012}.

In this work we have evaluated the polarimetric signals produced by stellar prominences in the \HeI\ 10830 and 5876 \AA\ triplets. The prominence has been characterized by a cloud model. A slab of a defined size is located at a certain height and latitude and it is forced to corotate with the star. We have the freedom to choose the prominence magnetic field strength and orientation, among other parameters such as the thermal broadening or the optical depth. The synthesis performed with \hazel\ allows us to derive the temporal evolution of the integrated Stokes parameters during a complete rotation of the prominence around the star. Our results show that the prominence produces an emission feature when it is located out of the stellar disk. This feature drifts across the profile according to the Doppler shift produced by the rotational velocity. When the prominence covers the stellar disk an absorption feature appears in the intensity profile. The scattering polarization produces clear signals in the linear polarization, which are visible in Stokes $Q$ and $U$ profiles. The presence of a magnetic field with a 10 G strength (close to that expected for actual stellar prominences) modifies the scattering event and changes the linear polarization measurements. The interpretation of these Stokes profile variations can potentially be used to infer coronal magnetic fields. When the magnetic field is directed along the LOS some circular polarization is generated through the Zeeman effect and it appears on the Stokes $V$ profile. However, its amplitude is very low.

The signals found for the particular cases explored in this work are subject to the choices made for several parameters. For example, in all the discussed examples we have set the optical depth to unity. An independent analysis was performed with a thicker prominence (not shown in the figures) and the results exhibit significant higher linear polarization signals. \citet{Dunstone+etal2006b} estimated the optical depth from several stellar prominences in H$\alpha$ and \CaIIH\ and retrieved average values of 22.4 and 15.6, respectively. This indicates that thick prominences are not uncommon. We would need a measurement of the optical depths at the wavelengths of the helium lines analized in this paper in order to refine our estimations.

Based on detailed analyses of instrument optical and detector performance with the MIRADAS data simulator, the signal-to-noise (S/N) expected for MIRADAS can be estimated as

\begin{equation}
\label{eq:geff}
\mathrm{S}/\mathrm{N}(T,m)=10\sqrt{T}10^{-\frac{m-18}{5}}f,
\end{equation}

\noindent where $T$ is the exposure time in hours, $m$ is magnitude of the star in the J-band, and $f$ is an efficiency
correction factor for the polarimetric modulation. The measurement of the four Stokes parameters requires a modulation (multiplexing) scheme that typically consists of $4$ exposures (or sometimes up to 12). These schemes roughly give $1/\sqrt{3}$ of the total number of photons to each Stokes parameter, so that the S/N is reduced by a factor $f\approx 1/\sqrt{3}$ \citep{delToroIniesta+Collados2000}. 

The estimated amplitude of Stokes $Q$ in the in \HeI\ 5876 \AA\ triplet is about 0.44\% of the intensity continuum, and Stokes $U$ signal is around 0.33\% (Figure \ref{fig:stokes_prom6}). The linear polarization signatures of the \HeI\ 10830 \AA\ triplet show a maximum amplitude around 0.6\% of the intensity contiuum (Figures \ref{fig:stokes_prom2}-\ref{fig:stokes_prom5}) for the magnetized prominences, while for the unmagnetized case the amplitude of Stokes $Q$ is about 1.6\% (Figure \ref{fig:stokes_prom1}). A reliable detection of the scattering polarization signals in the \HeI\ 10830 \AA\ triplet requires a S/N in linear polarization around 300. Assuming the implementation of efficient modulation schemes commonly used in solar telescopes, this S/N can be reached for a star with the J magnitude of AB Dor ($m=5.316$) with 82 seconds of exposure time. Unfortunately, AB Dor, the ideal candidate for this study, is not visible from the northern hemisphere where MIRADAS will be located. In the northern sky, stellar prominences have been detected in the G dwarf AP 149 \citep[located in the open cluster $\alpha$ Persei,][]{Barnes+etal2001}, in RE 1816+541 \citep{Eibe1998}, and in the M dwarf HK Aquarii \citep{Byrne+etal1996}. The later has a J magnitude of 7.979 and obtaining observations with S/N=300 would require 16 minutes exposures. Detecting linear polarization signals in this star would be challenging but possible. It must be noted that our estimation of the polarization signals was obtained assuming an optical depth unit in the \HeI\ lines, and it might be lower than the actual value. Thicker prominences will generate a much stronger signal that could be easily measured.

In this study, we have focused on the analysis of a simple isolated
prominence. However, cool stars can host complex prominence systems, with
several prominences simultaneously on the observable hemisphere \citep[\eg,][]{Dunstone+etal2006a}. In the ideal case, if two prominences are located at
different longitudes on the stellar disk, their individual Stokes profiles
could be analysed separately since they would appear at different Doppler
shifts. On the other hand, when prominences are close, their contribution
to the signal cannot be isolated. The measurement at each wavelength would
be the sum of the profiles produced by both prominences, and this addition
can be constructive if their magnetic fields are in the same direction or
the signal can be lost if their magnetic fields are perpendicular. The
presence of many prominences in the system will significantly complicate
the interpretation of the polarimetric signals. The integrated polarization
discussed in Sect. \ref{sect:integrated} could also be weak due to the cancellation of
signatures with opposite polarities, in this case independently of their
Doppler shift.

The \HeI\ 10830 and 5876 \AA\ triplets chosen for this work are among the
best candidate lines for probing chromospheric plasmas, since their
modeling and interpretation are relatively straightforward. Other visible
and near infrared lines could also be considered as possible tracers of
prominence magnetic fields, including the Balmer and Paschen series of
hydrogen or the \CaII\ infrared triplet. H$\alpha$ is the most common line for
the detection of stellar prominences, although its complex formation makes
it difficult to use it for plasma diagnostics. One could attempt to perform
multiline techniques, similar to the study of photospheric magnetic fields
using Zeeman-Doppler imaging, in order to increase the S/N. This approach
presents several difficulties. Contrary to the Zeeman effect, the profile
shapes of the scattering polarization signals strongly depends on specific
details of the transition. \citet{Belluzzi+LandiDeglInnocenti2009} classified the linear polarization signals from all lines in the second
solar spectrum into five classes based of the shape of their Q/I profile.
The Stokes Q profile of H$\alpha$ measured in solar prominences show a single
peak at the line center \citep{LopezAriste+etal2005}, and this is the most
common signature according to the examination from \citet{Belluzzi+LandiDeglInnocenti2009}. The number of chromospheric lines subject to be
included in a line-addition technique for stellar prominences is of the
order of a few tens. Assuming that all of them correspond to the same class
and can be coherently added, an increase of only a factor 3-5 can be
expected in the S/N.

For the magnetic field strength of the analyzed prominences (up to 10 G) the polarization in Stokes $V$ is too weak to allow a trustworthy detection with MIRADAS. Significantly stronger field strengths are unlikely at the large radial distances where stellar prominences are found. This way, the polarization produced by the Zeeman effect is not expected to be detected by current or near future infrastructures. The analysis and interpretation of the scattering polarization and the Hanle effect are the most promising approaches to attempt a direct evaluation of magnetic field in stellar prominences and stellar coronae. Our results show that this detection will be within reach of MIRADAS for the nearest rapidly rotating cool stars.


\section*{Acknowledgements}  
Financial support by the Spanish Ministry of Economy and Competitiveness
through projects AYA2014-60476-P and AYA2014-60833-P are gratefully acknowledged. 
AAR also acknowledges financial support through the Ram\'on y Cajal fellowship.
This research has made use of NASA's Astrophysics Data System Bibliographic Services.

\bsp	
\label{lastpage}
\end{document}